\documentclass[aps,prl,twocolumn,showpacs,10pt,superscriptaddress,preprintnumbers,nofootinbib]{revtex4-2}
\usepackage{amsmath,bm}
\allowdisplaybreaks[1]
\usepackage{physics}
\usepackage{graphicx}
\usepackage{amssymb}
\usepackage{xcolor}
\usepackage[colorlinks, linkcolor=blue!80!black, citecolor=green!50!black, urlcolor=magenta]{hyperref}
\usepackage[normalem]{ulem}
\usepackage{adjustbox}

\newcommand{\psec}[1]{\textbf{\emph{#1.}}}

\newcommand{\fig}[1]{Fig.~\ref{#1}}
\newcommand{\eq}[1]{Eq.~\eqref{#1}}

\newcommand{\refcite}[1]{Ref.~\cite{#1}}

\newcommand{\pp}[1]{\left(#1\right)}
\newcommand{\bb}[1]{\left[#1\right]}

\newcommand{\beq}[1][]{\begin{equation}\label{#1}}
\newcommand{\eeq}{\end{equation}}
\newcommand{\bse}[1][]{\begin{subequations}\label{#1}}
\newcommand{\ese}{\end{subequations}}
\newcommand{\nn}{\nonumber}

\renewcommand{\P}{\mathcal{P}}
\newcommand{\wt}[1]{\widetilde{#1}}

\newcommand{\M}{\mathcal{M}}
\newcommand{\Mt}{\wt{\mathcal{M}}}

\newcommand{\F}{\mathcal{F}}
\renewcommand{\H}{\mathcal{H}}

\def\GeV{{\rm GeV}}

\begin{document}

\preprint{
	{\vbox {			
		\hbox{\bf JLAB-THY-23-3828}
		\hbox{\bf MSUHEP-23-015}
}}}
\vspace*{0.2cm}

\title{Extraction of the Parton Momentum-Fraction Dependence of Generalized Parton Distributions from Exclusive Photoproduction}

\author{Jian-Wei Qiu}
\email{jqiu@jlab.org}
\affiliation{Theory Center, Jefferson Lab,
Newport News, Virginia 23606, USA}
\affiliation{Department of Physics, William \& Mary,
Williamsburg, Virginia 23187, USA}

\author{Zhite Yu}
\email{yuzhite@msu.edu}
\affiliation{Department of Physics and Astronomy, 
Michigan State University, East Lansing, Michigan 48824, USA}

\date{\today}

\begin{abstract}
The $x$ dependence of hadrons' generalized parton distributions (GPDs)
$\F(x,\xi,t)$ is the most difficult to extract from the existing known processes, 
while the $\xi$ and $t$ dependence are uniquely determined by the kinematics of the scattered hadron.
We study the single diffractive hard exclusive processes for extracting GPDs 
in the photoproduction. We demonstrate quantitatively the enhanced sensitivity on extracting the $x$ dependence 
of various GPDs from the photoproduction cross sections, as well as the asymmetries constructed from photon polarization 
and hadron spin that could be measured at JLab Hall D by GlueX Collaboration and future facilities.
\end{abstract}

\maketitle

\psec{Introduction}---
The generalized parton distributions (GPDs), $\F(x, \xi, t)$, 
provide rich information on the confined spatial distributions of quarks and gluons inside a bound hadron
(for reviews, see~\cite{Goeke:2001tz, Diehl:2003ny, Belitsky:2005qn, Boffi:2007yc}).
The Fourier transform of their $t$ dependence at the forward limit $\xi\to0$ provides
tomographic quark and gluon images of the hadron in its transverse plane
as functions of the active parton momentum fraction $x$~\cite{Burkardt:2000za, Burkardt:2002hr}.
The $x$ moments of GPDs are responsible for many emergent hadronic properties such as the hadron's 
mass~\cite{Ji:1994av, Ji:1995sv, Lorce:2017xzd, Metz:2020vxd} and spin~\cite{Ji:1996ek}, 
as well as its internal pressure and shear force~\cite{Polyakov:2002yz, Polyakov:2018zvc, Burkert:2018bqq}.

\begin{figure}[htbp]
\centering
	\begin{tabular}{cc}
		\includegraphics[trim={0.3cm 0 0.2cm 0}, clip, scale=0.57]{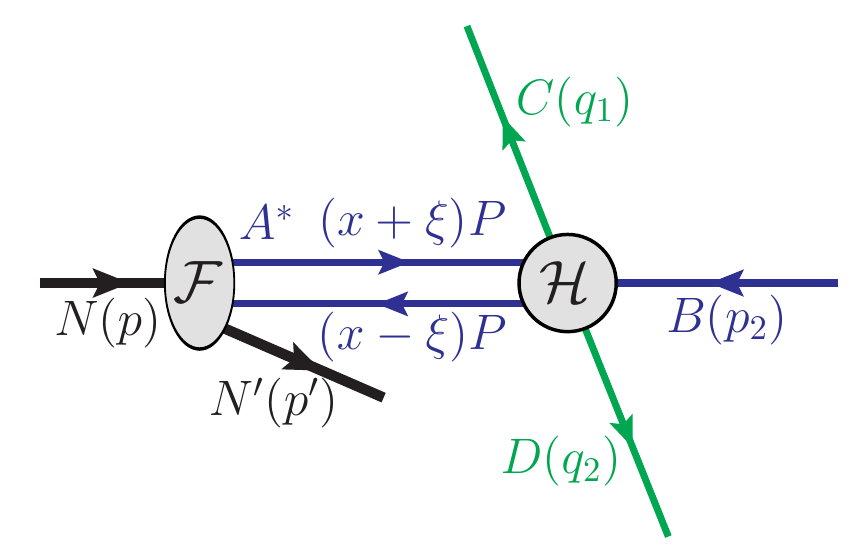} &
		\includegraphics[scale=0.57]{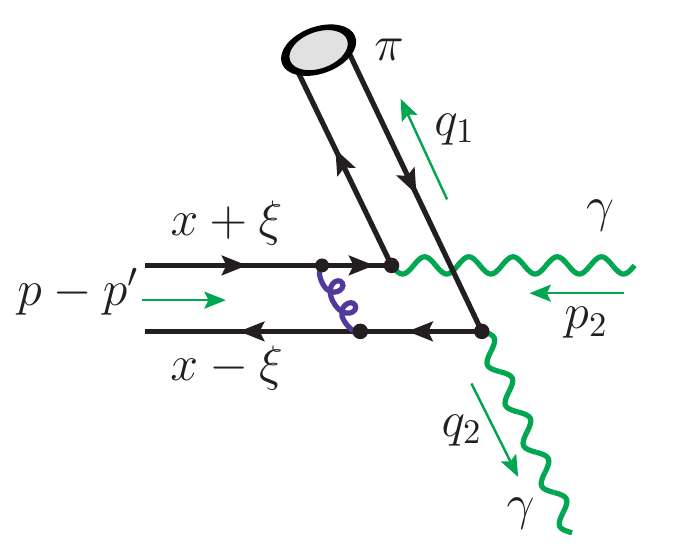}	\\
		(a) & (b) 
	\end{tabular}
\caption{(a) Sketch of $2\to 3$ SDHEP needed for extracting GPDs.  
(b) Sample diagram for the SDHEP in \eq{eq:sdhep}. }
\label{fig:sdhep}
\end{figure}

Experimental measurement of GPDs
requires a $2\to 3$ exclusive process at a minimum, 
as sketched in Fig.~\ref{fig:sdhep}(a), 
in which a hadron $N$ of momentum $p$ is scattered (or diffracted) to a hadron $N'$ of momentum $p'$
by exchanging a virtual two-parton state $A^*$ of momentum $\Delta\equiv p-p'$ and invariant mass $t = \Delta^2$, 
which undergoes a hard exclusive scattering with the colliding particle $B=\,$(lepton, photon, pion) of momentum $p_2$ 
to produce two back-to-back particles $C(q_1)$ and $D(q_2)$. 
To ensure the separation between the hard scattering $\H$ and the probed GPD $\F$, 
it is necessary to require the transverse momentum of the produced particles $C$ and $D$ 
to be much larger than the invariant mass of the exchange state $A^*$, $|q_{1T}|=|q_{2T}|\equiv q_T\gg \sqrt{|t|}$ 
(or equivalently, the hard collision time to be much shorter than the lifetime of the $A^*$) to 
suppress the quantum interference between the $\H$ and $\F$~\cite{Qiu:2022bpq,Qiu:2022pla}. 
We referred to such an exclusive process for extracting GPDs as the single diffractive hard exclusive process 
(SDHEP).  
By exchanging different $A^*$, SDHEP can probe different GPDs, 
${\cal F} \to H, \widetilde{H}, E, \widetilde{E}, \dots$, 
of different flavors~\cite{Goeke:2001tz, Diehl:2003ny, Belitsky:2005qn, Boffi:2007yc}.
A number of $2\to 3$ SDHEPs have been proposed for extracting 
GPDs~\cite{Ji:1996nm, Radyushkin:1997ki, Brodsky:1994kf, Frankfurt:1995jw, Berger:2001xd, Berger:2001zn, 
Kumano:2009he, ElBeiyad:2010pji, Pedrak:2017cpp, Siddikov:2022bku}, 
among which is the deeply virtual Compton scattering (DVCS)~\cite{Ji:1996nm, Radyushkin:1997ki},
corresponding to $B=C=\text{electron}$ and $D=\gamma$.
In addition, a few $2\to 4$ SDHEPs have also been proposed for extracting 
GPDs~\cite{Guidal:2002kt, Belitsky:2002tf, Belitsky:2003fj, Pedrak:2020mfm}.

Once the scattered hadron momentum $p'$ is measured, the $t$, $\xi\equiv \Delta^+/(2 P^+)$ with $P = (p + p')/2$,
and collision energy of the hard exclusive subprocess $(\Delta+p_2)^2 $ are fully determined.  
For an SDHEP to be sensitive to the $x$ dependence of GPDs, the remaining freedom of the hard subprocess $\H$, 
such as the $q_T$ (or the angle) of the produced particle $C$ or $D$, needs to be {\it entangled} with $x$, 
which is proportional to the relative momentum of the two exchanged partons~\cite{Qiu:2022pla}.  
For the DVCS, the exchange state $A^*(\Delta)$ in Fig.~\ref{fig:sdhep} can be a virtual photon 
for the Bethe-Heitler process, a $q\bar{q}$ pair for quark GPDs, and a pair of gluons for gluon GPDs,
if we neglect terms further suppressed by powers of $Q^2 = -(p_2 - q_1)^2$.  
Since the relative momentum of the two exchanged partons is decoupled from external variation of $Q^2$ 
at leading order, the measured DVCS cross sections probe GPDs through their ``moments,'' 
like $\int dx\, \F(x,\xi,t)/(x-\xi)$~\cite{Qiu:2022pla}, which makes it very difficult
to extract the full $x$ dependence of GPDs~\cite{Bertone:2021yyz}.
Although QCD evolution of GPDs could introduce some sensitivity to the $x$ dependence~\cite{Moffat:2023svr}, 
the event rate drops very quickly when $Q^2$ increases. 

In this Letter, we study the sensitivity in extracting GPDs from exclusive photoproduction~\cite{Qiu:2022pla, Duplancic:2018bum,Duplancic:2022ffo},
\beq[eq:sdhep]
	N(p) + \gamma(p_2) \to N'(p') + \pi(q_1) +  \gamma(q_2).
\eeq
The corresponding QCD factorization was justified in Ref.~\cite{Qiu:2022pla} by treating this process as a crossing process of 
the exclusive diphoton production in diffractive pion-nucleon collisions~\cite{Qiu:2022bpq}.
We calculate the leading-order (LO) short-distance hard parts 
and find that the transverse momentum (or the polar angle $\theta$)
of the final-state pion is clearly entangled with the relative momentum of the two exchanged partons.  
Variation of observed $q_T$ can provide enhanced sensitivity to the $x$ dependence of GPDs. 
With the crossing kinematics, this process provides more enhanced $x$ sensitivity in the 
ERBL-region of GPDs defined to be the region where $|x| \leq |\xi|$,
and the diphoton production in pion-nucleon scattering is more sensitive to the DGLAP-region ($|x| > |\xi|$)~\cite{Qiu:2022bpq}.
In addition, with the well-controlled polarization of the initial-state photon beam at JLab Hall D~\cite{GlueX:2017zoo} 
and polarized hadron targets, we introduce asymmetries of cross sections constructed from 
the photon beam polarization and target spin and 
demonstrate quantitatively the enhanced capability of extracting various GPDs and their $x$ dependence from 
measurements at JLab Hall D and future facilities.

\begin{figure}[htbp]
	\centering
	\includegraphics[scale=0.44]{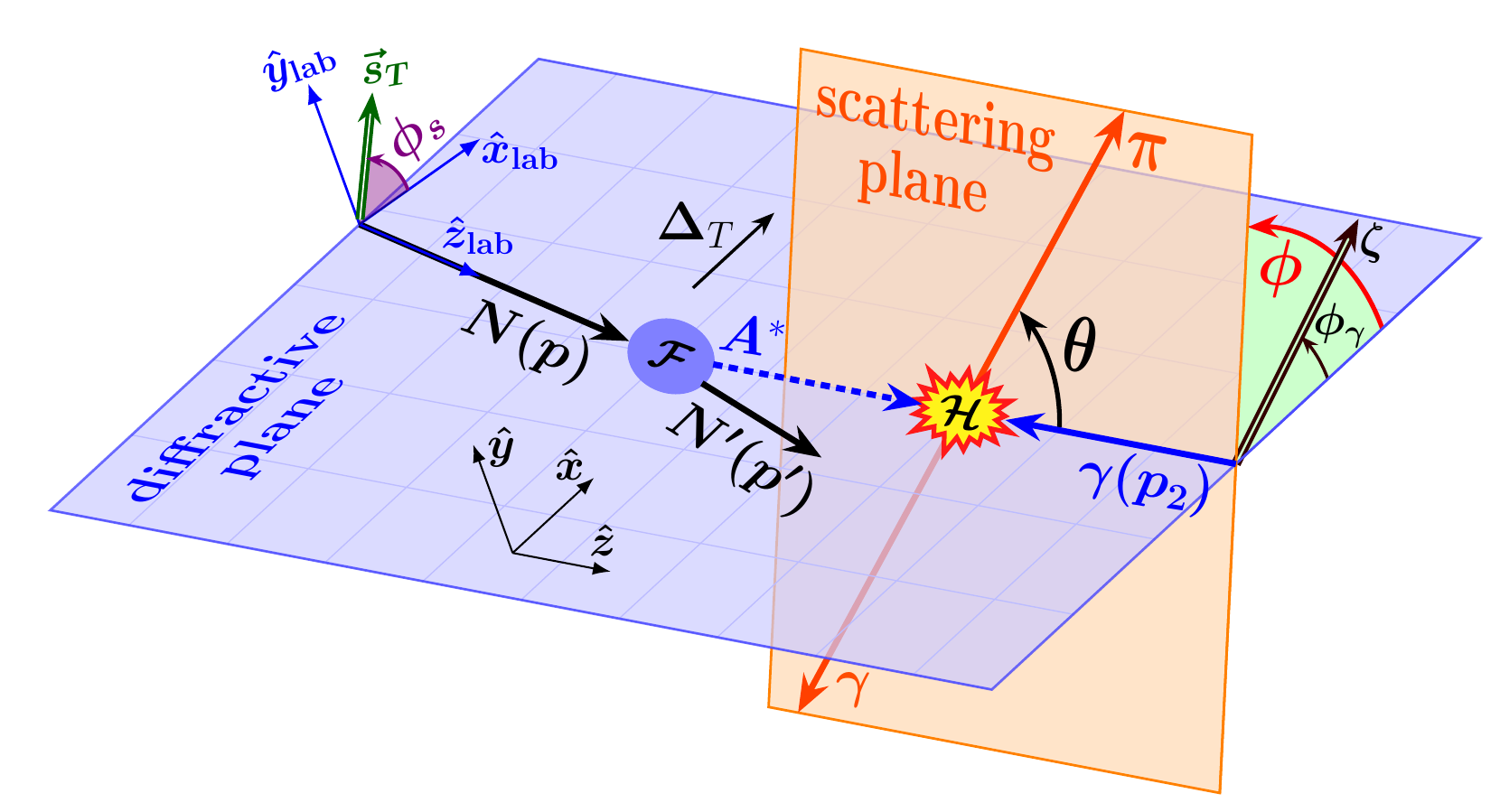}
	\caption{Frames for the process in \eq{eq:sdhep}. 
		The vectors of $\bm{s}_T$ and $\zeta$ refer to the transverse spin and linear polarization of 
		the colliding nucleon $N$ and photon $\gamma$, respectively.
	}
	\label{fig:frame}
\end{figure}

\psec{Kinematics and observables}---
In \fig{fig:frame}, we describe the kinematics of the SDHEP in \eq{eq:sdhep} 
in terms of two frames and two planes.
The {\it Lab frame} is chosen to be the center-of-mass (c.m.) frame 
of the colliding hadron $N(p)$ and photon $\gamma(p_2)$ with 
the $\hat{z}_{\rm lab}$ along the momentum $p$, and $\hat{x}_{\rm lab}$
on the $N \to N'$ {\it diffractive plane} defined by the momentum $p$ and $p'$.  
The {\it SDHEP frame} is the c.m.~frame of the final-state $\pi$-$\gamma$ pair, 
which is the same as the c.m.~of the hard scattering subprocess, with 
$\hat{z}$ along the momentum $\Delta$ of $A^*$, while the initial-state photon travels along the $-\hat{z}$ direction 
and $\hat{x}$ lies on the diffractive plane.  
The $\hat{z}$ and the observed $\pi$ momentum $q_1$ define the {\it scattering plane},
and the angles $(\theta, \phi)$ define the direction of the observed $\pi$ in the SDHEP frame.
Choosing the ($\hat{x}_{\rm lab},\hat{z}_{\rm lab}$) and ($\hat{x},\hat{z}$) of these two frames on the same diffractive plane
makes the Lorentz transformation between them simpler.

The SDHEP frame in Fig.~\ref{fig:frame} is very similar to the Breit frame for describing
the lepton-hadron semi-inclusive deep inelastic scattering (SIDIS) in the Trento convention~\cite{Bacchetta:2004jz} 
if one corresponds the hadron $N$ and $N'$ to the colliding electron and scattered electron in SIDIS, respectively, 
and the diffractive plane and scattering plane to the leptonic plane and hadronic plane in SIDIS, respectively.  
But, unlike the highly virtual photon exchanged between the colliding lepton and hadron in SIDIS, 
the $A^*$ is a ``long-lived'' state with a low enough virtuality.

Both the colliding photon and hadron target at JLab Hall D can be polarized longitudinally. 
In addition, the photon can have linear polarization $\zeta$ and the 
hadron can have a transverse spin $s_T$,
defined by the azimuthal angles, $\phi_{\gamma}$ and $\phi_s$ in the Lab frame, respectively.

Having a pion in the final state eliminates the contribution from $A^*$ being a virtual photon due to charge parity, so that
the leading contribution to the SDHEP in \eq{eq:sdhep} is from channels with $A^*$ being a collinear parton pair.
The corresponding scattering amplitude can be factorized into 
GPDs for the hadron transition $N\to N'$, a distribution amplitude (DA) for the formation of the final-state pion, and 
perturbatively calculable coefficients~\cite{Qiu:2022pla}
\begin{align}\label{eq:factorize}
	&\M_{ N \gamma_{\lambda} \to N' \pi \gamma_{\lambda'} }^{[\F, \wt{\F}]}
	= \sum_{f,f'} \int_{-1}^{1} dx \int_0^1 dz \, \bar{D}_{f'/\pi}(z)  \\
	&\;\; \times	\bb{ 
			\F_{NN'}^f(x, \xi, t) \, \widetilde{C}^{ff'}_{\lambda\lambda'}(x, z)
			  + \wt{\F}_{NN'}^f(x, \xi, t) \, C^{ff'}_{\lambda\lambda'}(x, z)
		}, \nn
\end{align}
where $f=[q\bar{q}]$ and $[gg]$ for quark and gluon GPDs, respectively, if $N'=N$, 
or $f=[q\bar{q}']$ for transition GPDs with $N \neq N'$, and correspondingly, 
$f'=[q\bar{q}]$ or $[q\bar{q}']$ for the pion DA $\bar{D}_{f'/\pi}$.  
The hard coefficients $C^{ff'}_{\lambda\lambda'}$ and $\widetilde{C}^{ff'}_{\lambda\lambda'}$ 
are helicity amplitudes for the photon scattering off a collinear on-shell parton pair $f$ 
with $\lambda$ and $\lambda'$ denoting the photon helicities in the SDHEP frame.
Under the parity invariance, they can be reduced to 
two helicity-conserving amplitudes ($C_{+}$, $\wt{C}_{+}$) 
and two helicity-flipping ones ($C_{-}$, $\wt{C}_{-}$).
Their explicit forms are collected in Supplemental Material~\cite{sm}.
The correction to the factorization in Eq.~(\ref{eq:factorize}) is suppressed by powers of $|t|/q_T^2 \ll 1$.

The differential cross section for the SDHEP in \eq{eq:sdhep} is
\begin{align}\label{eq:xsec}
	&\frac{d\sigma}{d |t| \, d\xi \, d\cos\theta \, d\phi}
	= \frac{1}{2\pi} \frac{d\sigma}{d |t| \, d\xi \, d\cos\theta }
		\cdot
		\big[ 1 +  \lambda_N \lambda_{\gamma} \, A_{LL} \nn\\
	& \hspace{0.6em} 
				+ \zeta \, A_{UT} \cos2(\phi - \phi_{\gamma})
				+ \lambda_N \, \zeta A_{LT} \sin2(\phi - \phi_{\gamma})
		\big],
\end{align}
where $\lambda_N$ and $\lambda_{\gamma}$ are the net helicities of the initial-state nucleon and photon, respectively.  
In \eq{eq:xsec}, we introduced the unpolarized differential cross section,
\begin{align}\label{eq:unpol-xsec}
	\frac{d\sigma}{d |t| \, d\xi \, d\cos\theta }
	= \pi (\alpha_e \alpha_s)^2 \pp{ \frac{C_F}{N_c} }^2 \frac{1 - \xi^2}{\xi^2 s^3}\, \Sigma_{UU},
\end{align}
with $\Sigma_{UU}$ and the polarization asymmetries given by
\begin{align}\label{eq:asymmetries}
	\Sigma_{UU}
	&= | \M_{+}^{[\wt{H}]} |^2
		+ | \M_{-}^{[\wt{H}]} |^2
		+ | \wt{\M}_{+}^{[H]} |^2 
		+ | \wt{\M}_{-}^{[H]} |^2,	\nn\\
	A_{LL}
	&= 2 \, \Sigma_{UU}^{-1} \Re\bb{
					\M_{+}^{[\wt{H}]} \, \wt{\M}_{+}^{[H] *}
					+ \M_{-}^{[\wt{H}]} \, \wt{\M}_{-}^{[H] *}
		} , \nn\\
	A_{UT}
	&= 2 \, \Sigma_{UU}^{-1} \Re\bb{
					\wt{\M}_{+}^{[H]} \, \wt{\M}_{-}^{[H] *}						
					- \M_{+}^{[\wt{H}]}  \, \M_{-}^{[\wt{H}] *}
		} , \nn\\
	A_{LT}
	&= 2 \, \Sigma_{UU}^{-1} \Im\bb{
					\M_{+}^{[\wt{H}]} \, \wt{\M}_{-}^{[H] *}		
					+ \M_{-}^{[\wt{H}]} \, 	\wt{\M}_{+}^{[H] *}	
		} ,
\end{align}
whose two subscripts are for hadron spin and photon polarization, respectively, with 
$U$ for ``unpolarized,'' $L$ for ``longitudinal polarized,'' and $T$ for ``linearly polarized photon,'' 
leaving the situation of transversely polarized hadron to a future publication.  
The helicity amplitudes $\M_{\pm}^{[\wt{H}]}$ and $\wt{\M}_{\pm}^{[H]}$
in \eq{eq:asymmetries} are given as 
convolutions of GPD $\wt{H}$ and $H$, respectively, and for example,
\begin{equation}
\M_{\pm}^{[\wt{H}]} 
	= \int_{-1}^1 dx \int_0^1 dz \, \wt{H}(x, \xi, t) \, \bar{D}(z) \, C_{\pm}(x, z; \theta)
\end{equation}
with corresponding hard coefficients given in Supplemental Material~\cite{sm}.
In this Letter, we focus on contributions from quark GPDs and 
leave the contribution of gluon GPDs to a future publication.  

\psec{Enhanced $x$-sensitivity}---
While GPDs' $t$ and $\xi$ dependence can be directly measured,
their $x$ dependence (as well as the $z$ dependence of DA) is only probed 
via convolutions as in \eq{eq:factorize}. 
As explained in \refcite{Qiu:2022pla},
the LO hard coefficient $C$ for almost all known processes for extracting GPDs has its 
$x$ dependence decoupled from the measured hard scale(s), e.g., for DVCS~\cite{Ji:1996nm, Radyushkin:1997ki}, 
\beq[eq:hard-coef-factorize]
	C_{\rm DVCS}^{\rm LO}(x, \xi; x_B, Q^2) = \frac{1}{x\pm\xi\mp i\varepsilon}\, C(x_B, Q^2)\, .
\eeq
Consequently, experimental variation of the probing scale of these processes,
such as $(x_B,Q^2)$ here,
has little influence on the $x$ convolution of GPDs. 
Since the unpinched $x$ poles in \eq{eq:hard-coef-factorize} are only localized at $\pm\xi$, experimental measurements of DVCS 
may only constrain the diagonal values of GPDs $\F(\xi, \xi, t)$ through the imaginary parts
and the limited ``moments,''
\beq[eq:GPD-moments]
	 \F_0(\xi, t) = \P \int_{-1}^1 dx \frac{\F(x, \xi, t)}{x - \xi},
\eeq
with $\P$ indicating principle-value integration.
Such lack of sensitivity to the full $x$ dependence of GPDs
is also true for other known processes, 
including the deeply virtual meson production (DVMP)~\cite{Brodsky:1994kf, Frankfurt:1995jw}, 
photoproduction of lepton~\cite{Berger:2001xd} 
or photon pair~\cite{Pedrak:2017cpp, Grocholski:2021man, Grocholski:2022rqj}, 
and the exclusive Drell-Yan process~\cite{Berger:2001zn}. 

Having only the moment sensitivity is far from enough to map out the $x$ distribution of GPDs. 
One can easily construct null solutions to \eq{eq:GPD-moments} that give zero to the 
moments, diagonal values, and forward limit~\cite{Bertone:2021yyz},
at which $(\xi, \, t) \to 0$ and GPDs reduce to parton distribution functions.
Such solutions are termed shadow GPDs, 
which are invisible to processes that only possess moment-type sensitivity.
Although QCD evolution of GPDs in response to the variation of the probing scale might 
help with this situation~\cite{Freund:1999xf}, the nature of logarithmic high-order contribution makes 
the improvement numerically not appreciable~\cite{Bertone:2021yyz} 
unless one goes to a sufficiently high scale~\cite{Moffat:2023svr} 
where the exclusive cross section itself diminishes, 
making it difficult to reach the desired precision.

In contrast, the hard coefficients for the SDHEP in \eq{eq:sdhep}, as collected in Supplemental Material~\cite{sm},  
have not only terms in which the $x$ dependence is decoupled from the external hard scale $q_T$ 
(or equivalently, the polar angle $\theta$) of the observed pion in the SDHEP frame, like that in \eq{eq:hard-coef-factorize},  
but also terms in which the $x$ dependence cannot be factorized as in \eq{eq:hard-coef-factorize} and 
is entangled with the observed $q_T$ (or $\theta$).  More precisely, 
the helicity-conserving hard coefficients $C_+$ and $\wt{C}_+$ 
contain terms proportional to the product of electric charge of quark and antiquark 
$e_1 e_2$, in which the external observable $\theta$ is entangled with the partons'
momentum fractions $x$ and $z$.  Their convolutions with GPD $H$ and $\wt{H}$ lead to 
the following type of integrals:
\beq[eq:special-integral]
	\int_{-1}^1 dx \frac{(H^+, \wt{H}^+)(x, \xi, t)}{x - x_p(\xi, z, \theta) + i \epsilon}\,  ,
\eeq
with the $x$ pole away from $\pm \xi$ and entangled with the externally measured $\theta$ in the form
\beq[eq:g-pole]
	x_p(\xi, z, \theta) = \xi \cdot \bb{ \frac{\cos^2(\theta / 2) (1 - z) - z}{\cos^2(\theta / 2) (1 - z) + z} }.
\eeq
Such contribution arises from Feynman diagrams with the two photons attached to two different fermion lines, 
like the one in Fig.~\ref{fig:sdhep}(b), so that the momentum flow through the short-distance gluon contains 
both $x$ dependence from the GPD (and $z$ dependence from DA)
and $q_T$ (or $\theta$) dependence.
This special gluon propagator is responsible for the $x_p$ form in \eq{eq:g-pole}.  
Such entanglement provides {\it enhanced sensitivity} to the $x$ dependence of GPDs from the experimentally 
measured  $q_T$ or $\theta$ distribution.  
With $z$ going from $0$ to $1$, $x_p$ in \eq{eq:g-pole} goes from $\xi$ to $-\xi$,
scanning through the whole ERBL region of GPDs.  
This is complementary to the high-$q_T$ diphoton production in single diffractive pion-nucleon scattering, 
which scans through the whole DGLAP region of GPDs~\cite{Qiu:2022bpq}.

The four helicity amplitudes $\M_{\pm}^{[\wt{H}]}$ and $\Mt_{\pm}^{[H]}$
cannot be distinguished by considering only the unpolarized differential cross section in \eq{eq:unpol-xsec}, from which
the two amplitudes $\M_+^{[\wt{H}]}$ and $\Mt_+^{[H]}$ with enhanced $x$ sensitivity cannot be distinguished.
Fortunately, with the capability of polarizing both the photon beam and hadron target at JLab, 
various polarization asymmetries can be constructed as shown in \eq{eq:asymmetries}.
The single spin asymmetry, $A_{UT}$, mixes the helicity-conserving and flipping amplitudes, 
and then depends more on the amplitudes with enhanced $x$ sensitivity, especially on their absolute signs.
The double spin asymmetries, $A_{LL}$ and $A_{LT}$, provide different combinations of the GPD $H$ and $\wt{H}$.
In particular, $A_{LT}$ is given by the imaginary parts of the amplitudes, which probe the GPD values in the ERBL region
due to the special $x$ pole at $x_p(\xi, z, \theta)$ in \eq{eq:g-pole}. 

The unpolarized cross section plus three asymmetries in \eq{eq:asymmetries}  
can provide good information to disentangle the GPDs $H$ and $\wt{H}$.
If the hadron can also be transversely polarized, the associated asymmetry can provide new
information to add constraints on the GPD $E$ and $\wt{E}$, which is beyond the scope of this Letter.

\begin{figure}[htbp]
	\centering
	\includegraphics[scale=0.38]{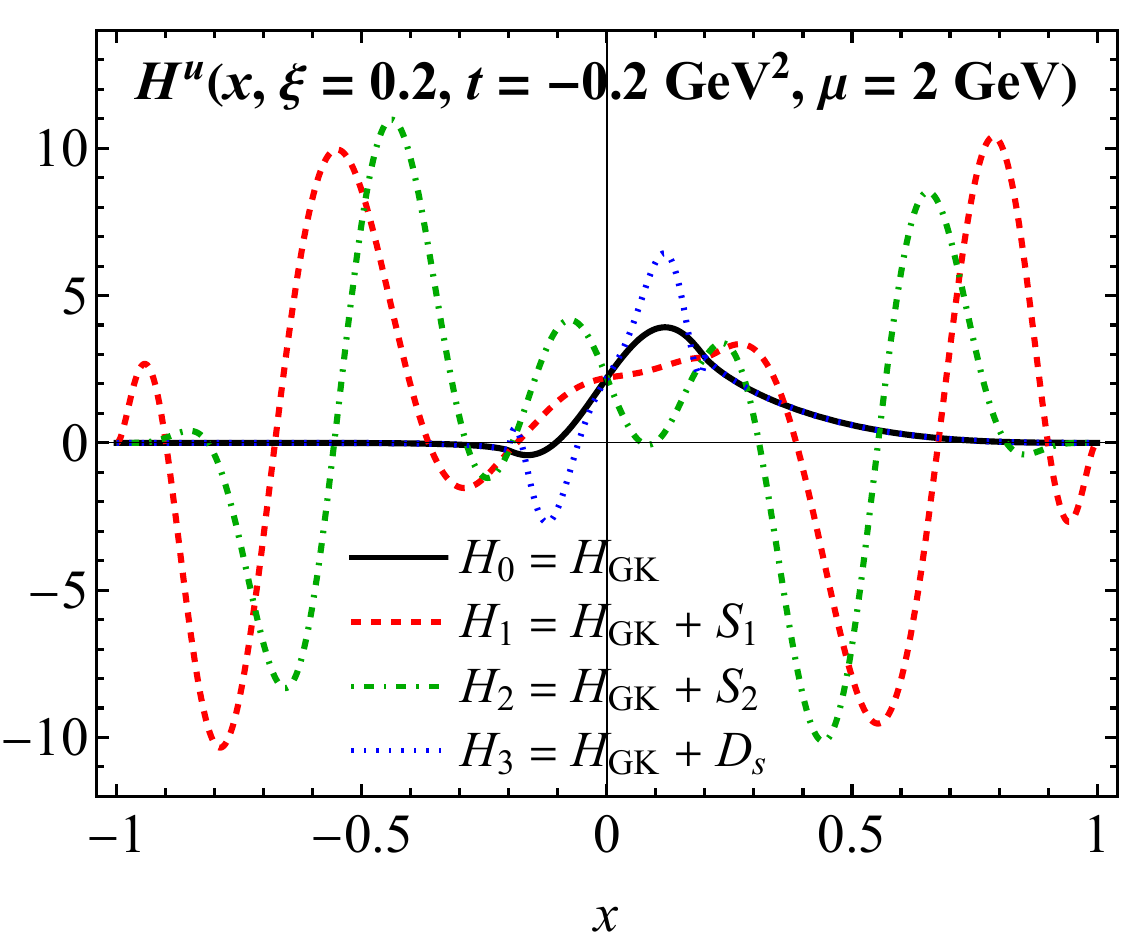}
	\includegraphics[scale=0.38]{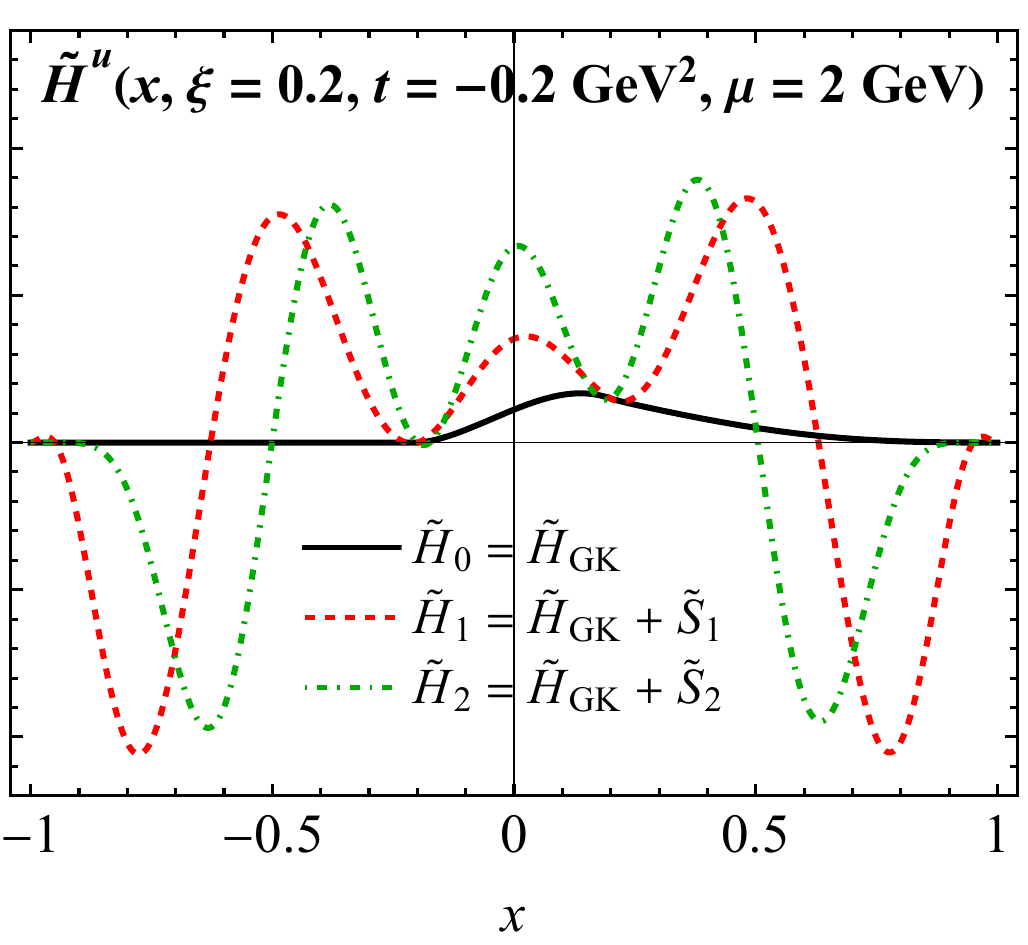} 
	\vspace{-3mm}
	\caption{Choices of the $u$-quark GPD models at $t = -0.2~\GeV^2$ and $\xi = 0.2$, by adding shadow GPDs to the GK model.}
	\label{fig:gpds}
\end{figure}

\psec{Numerical results}---
The CEBAF at JLab is capable of delivering intense polarized photon beam to its Hall D 
to study the SDHEP in \eq{eq:sdhep} on various hadron targets, which can also be polarized.
We evaluate the production rate and various asymmetries in \eq{eq:asymmetries} to 
demonstrate the enhanced $x$ sensitivity on extracting GPDs. 
We take the Goloskokov-Kroll (GK) model~\cite{Goloskokov:2005sd, Goloskokov:2007nt, Goloskokov:2009ia, Kroll:2012sm} 
as the reference GPD for $H$ and $\wt{H}$, referred to as $H_0$ and $\wt{H}_0$, respectively.  
As shown in \fig{fig:gpds}, we construct additional GPDs $H_i$ and $\wt{H}_i$ with different $x$ dependence from 
modifying the reference $u$-quark GPD by adding various shadow GPDs, 
$S_i(x, \xi)$ or $\wt{S}_i(x, \xi)$, or a shadow $D$-term $D_s(x/\xi)$, 
which are constructed (in Supplemental Material~\cite{sm}) to give zero contribution to the GPD's forward limit 
and moment in \eq{eq:GPD-moments}.
We fix the pion DA to be its asymptotic form~\cite{Lepage:1980fj}.
In order to focus on the $x$ sensitivity from the $q_T$ (or $\theta$) distribution of 
this particular process, we neglect evolution effects of GPDs and fix both renormalization and factorization scales at $2~\GeV$.

\begin{figure}[htbp]
	\centering
	\includegraphics[scale=0.52]{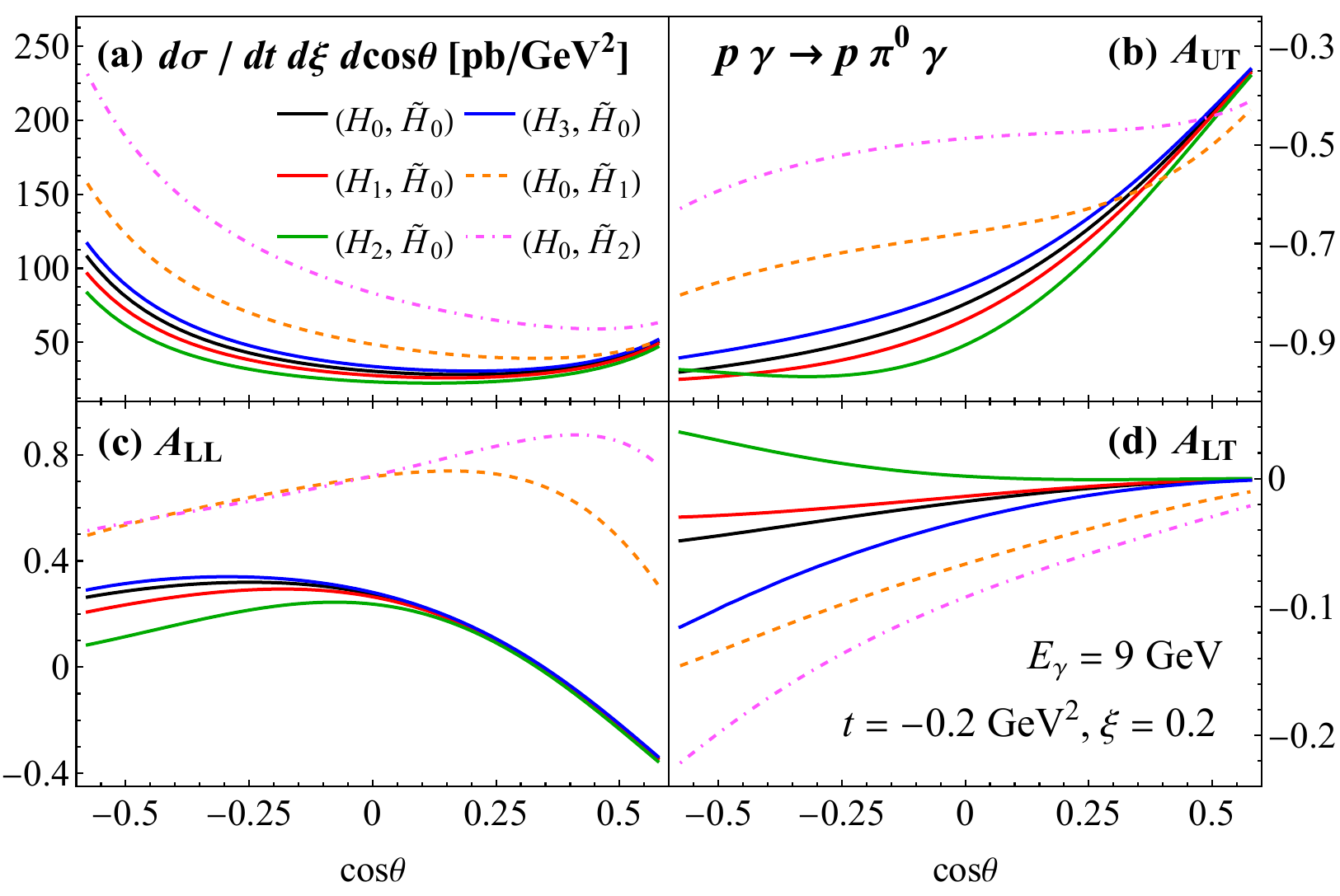}
	\vspace{-3mm}
	\caption{Unpolarized rate (a) and polarization asymmetries (b)--(d) 
	as functions of $\cos\theta$ at $(t, \xi) = (-0.2~\GeV^2, 0.2)$, 
	using different GPD sets as given in \fig{fig:gpds}.
	}
	\label{fig:pi0-distributions}
\end{figure}

In \fig{fig:pi0-distributions}, we show the unpolarized differential cross section in \eq{eq:unpol-xsec} together with 
the various asymmetries in \eq{eq:asymmetries} for $\pi^0$ production as a function of its polar angle $\theta$ 
in the SDHEP frame at $E_{\gamma} = 9~\GeV$.  
Since the $\cos\theta$ dependence of the hard coefficients $C_-$ and $\wt{C}_-$ is multiplicative to their
$x$ or $z$ dependence, they are not visible to the shadow GPDs.
On the other hand, the $\cos\theta$ dependence of $C_+$ and $\wt{C}_+$ is entangled with their 
$x$ and $z$ dependence,
and therefore, GPDs with different $x$ dependence lead to different rate and asymmetries.
In particular, the $A_{LT}$ is sensitive to the imaginary parts of the amplitudes, which are generated in
the ERBL region, and has better sensitivity to the shadow $D$-term than the other three observables 
as shown in \fig{fig:pi0-distributions}.
In general, the oscillation of shadow GPDs in the DGLAP region causes a big cancellation in 
their contribution to the amplitudes, 
while the sensitivity is more positively correlated with the GPD magnitude in the ERBL region.
The shadow $\wt{S}_i$ associated with the $x$ dependence of the polarized GPD $\wt{H}$ 
gives bigger contribution to the amplitude $\M_+^{[\wt{S}]}$ than $S_i$ to $\Mt_+^{[S]}$ 
due to charge symmetry property, so it can be better probed.

\begin{figure}[htbp]
	\centering
	\includegraphics[scale=0.52]{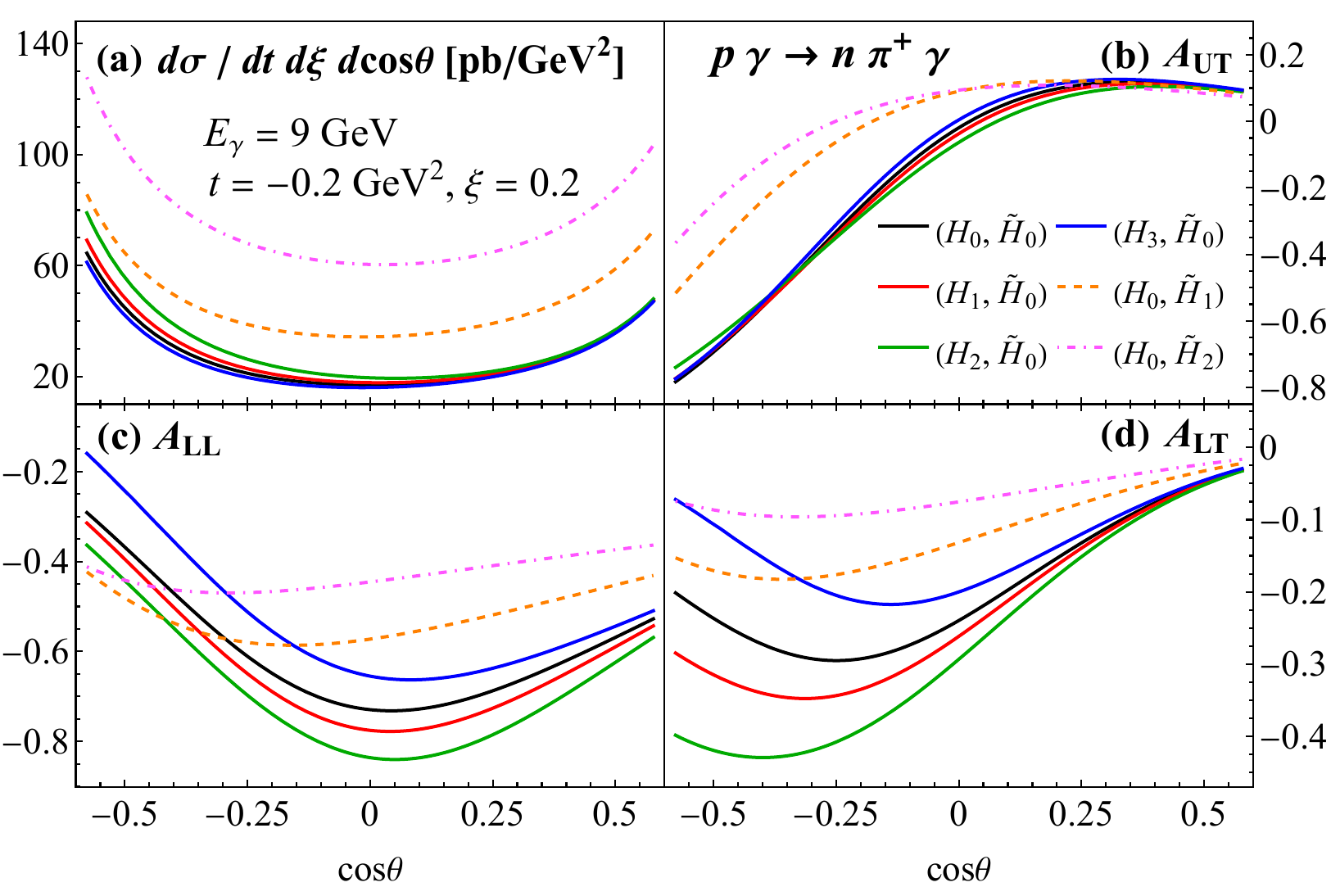}
	\vspace{-3mm}
	\caption{Same as \fig{fig:pi0-distributions}, but for the $p \gamma \to n \pi^+ \gamma$ process.}
	\label{fig:pi+distributions}
\end{figure}

For the neutral pion production, we can eliminate terms proportional to 
$(e_1-e_2)^2$ or $(e_1^2-e_2^2)$ in the hard coefficients since $e_1 = e_2$,
which effectively removes a good number of moment-type terms, 
giving the maximum amount of entanglement and the most 
sensitivity to GPDs' $x$ dependence.
In \fig{fig:pi+distributions}, we present the same study 
for the $p \gamma \to n \pi^+ \gamma$ process. With different flavor combination, 
it provides different $x$ sensitivity.  
The $n \gamma \to p \pi^- \gamma$ process gives a similar result, 
but with a smaller production rate.
As demonstrated in Figs.~\ref{fig:pi0-distributions} and \ref{fig:pi+distributions}, 
both the production rate and asymmetries are sizable and measurable, 
making the SDHEP in \eq{eq:sdhep} uniquely different from DVCS and others 
in terms of its enhanced sensitivity for extracting the $x$ dependence of GPDs.

\psec{Summary and outlook}---
Extracting the full $x$ dependence of GPDs is very important not only 
for probing the tomographic partonic images of hadrons, but also 
for predicting and understanding the emergent hadron properties 
in terms of various $x$ moments of GPDs.  
The fact that the most
known processes for extracting GPDs, including DVCS and DVMP, 
have only moment-type sensitivity makes it very difficult
to pin down the full $x$ dependence of GPDs and their flavor dependence
due to the possibility of having an infinite number of shadow GPDs 
which are hardly visible to these processes.  

In this Letter, we demonstrated quantitatively that the SDHEP in \eq{eq:sdhep} 
is not only accessible by JLab Hall D but also capable of providing 
much enhanced sensitivity to the $x$ dependence of GPDs, as well as
the potential to probe the flavor dependence of GPDs from production rates
and asymmetries involving various mesons.  This is possible because this process has 
the entanglement of the $x$ flow of GPDs with the externally observed hard scale~\cite{Qiu:2022pla, Qiu:2022bpq},
which is a critical criterion for searching for new physical processes to help extract the $x$ dependence of GPDs.
Since multiple GPDs could contribute to the same observables through convolutions
of their $x$ dependence, extracting GPDs from data is a challenging inverse problem. 
A global analysis of multiple processes is necessary for extracting these 
nonperturbative and universal GPDs from which we can picture the spatial distribution 
of the probability densities to find quarks and gluons inside a bound hadron.

With the full knowledge of the $x$, $\xi$ and $t$ dependence of GPDs, we would be able to 
not only address how partonic dynamics impacts the emergent hadronic properties, 
but also provide quantitative answers to profound questions, including 
what the proton radius is in terms of its transverse spatial distribution of quarks, $r_q(x)$, or
gluons, $r_g(x)$,
how such radii compare with its electromagnetic charge radius, 
and how far from the center of the proton the quarks and gluons could still be found.

\textbf{\emph{Acknowledgments.}}---We thank N.~Sato, J.~Stevens and M.~Strikman for helpful discussions and communications. 
This work is supported in part by the U.S. Department of Energy (DOE) Contract No.~DE-AC05-06OR23177, 
under which Jefferson Science Associates, LLC operates Jefferson Lab. 
The work of Z.Y. at MSU is partially supported by the U.S.~National Science Foundation under Grant No.~PHY-2013791, 
and the fund from the Wu-Ki Tung endowed chair in particle physics.

\bibliographystyle{apsrev}
\bibliography{reference}

\end{document}



\title{Supplemental material for ``Extraction of the Parton Momentum-Fraction Dependence of Generalized Parton Distributions from Exclusive Photoproduction''}

\author{Jian-Wei Qiu}
\email{jqiu@jlab.org}
\affiliation{Theory Center, Jefferson Lab,
Newport News, Virginia 23606, USA}
\affiliation{Department of Physics, William \& Mary,
Williamsburg, Virginia 23187, USA}

\author{Zhite Yu}
\email{yuzhite@msu.edu}
\affiliation{Department of Physics and Astronomy, 
Michigan State University, East Lansing, Michigan 48824, USA}

\date{\today}
\maketitle

\onecolumngrid

\makeatletter
\renewcommand\p@subsection{}
\makeatother

\setcounter{equation}{1000}
\setcounter{figure}{1000}
\setcounter{table}{1000}

\renewcommand{\theequation}{S\the\numexpr\value{equation}-1000\relax}
\renewcommand{\thefigure}{S\the\numexpr\value{figure}-1000\relax}
\renewcommand{\thetable}{S\the\numexpr\value{table}-1000\relax}

\vspace{-0.5in}
\subsection{A:\ Hard coefficients for the photon-proton scattering process}
\label{app:hard_coefs}
\vspace{-0.1in}

\begin{figure}[htbp]
	\centering
	\begin{tabular}{cccc}
		\includegraphics[scale=0.55]{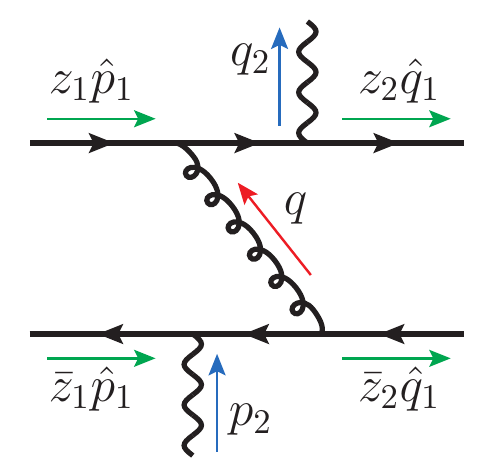} &
		\includegraphics[scale=0.55]{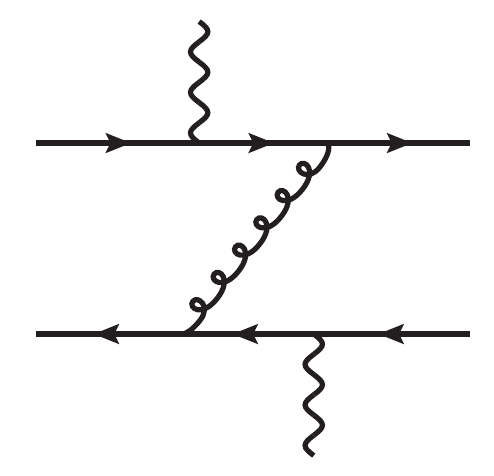} &
		\includegraphics[scale=0.55]{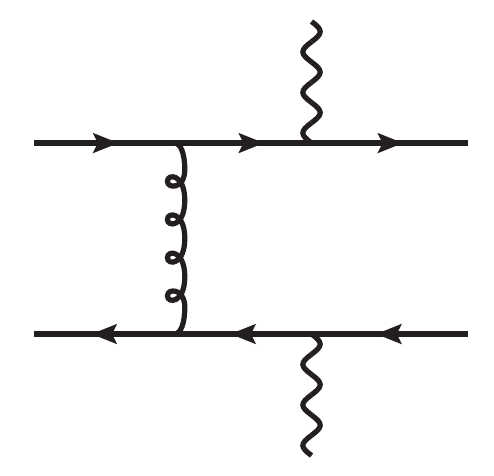} &
		\includegraphics[scale=0.55]{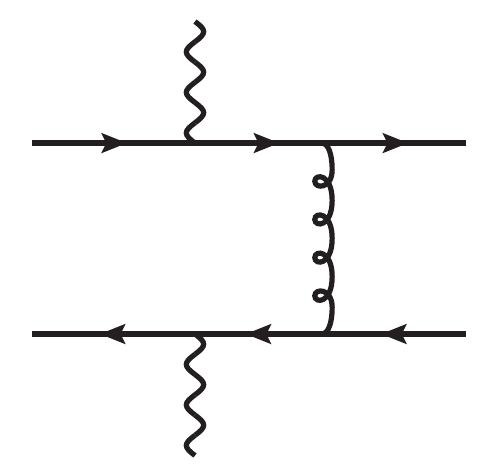} \\
		(A1) & (A2) & (A3) & (A4)
	\end{tabular}
	\begin{tabular}{cccccc}
		\includegraphics[scale=0.55]{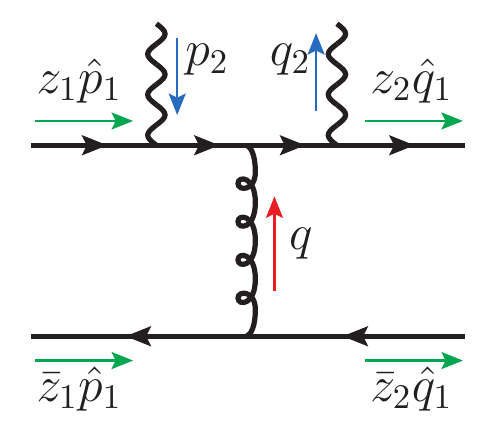} &
		\includegraphics[trim={0 -0.68cm 0 0}, clip, scale=0.55]{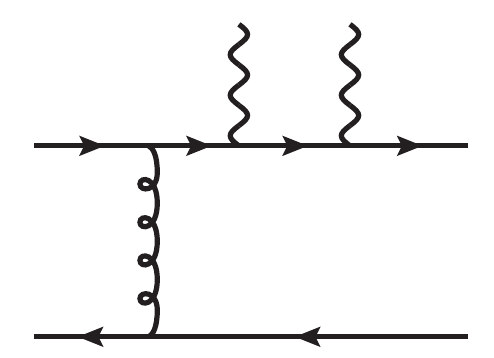} &
		\includegraphics[trim={0 -0.68cm 0 0}, clip, scale=0.55]{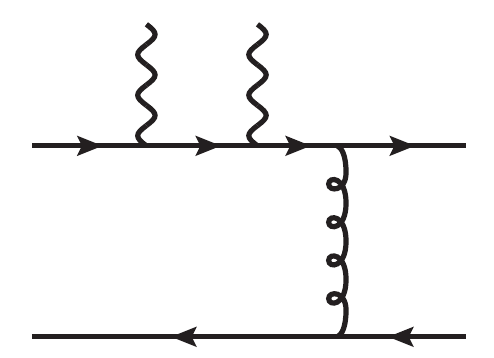} &
		\includegraphics[scale=0.55]{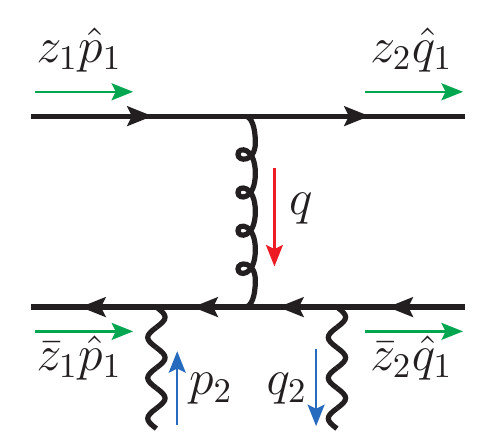} &
		\includegraphics[trim={0 0 0 -0.68cm}, clip, scale=0.55]{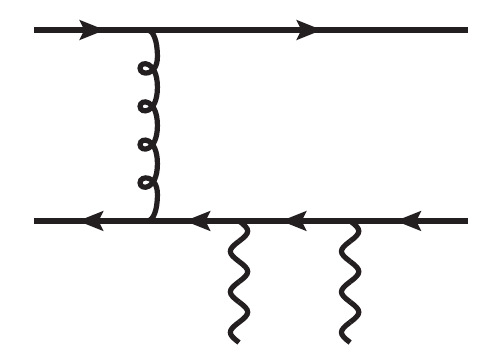} &
		\includegraphics[trim={0 0 0 -0.68cm}, clip, scale=0.55]{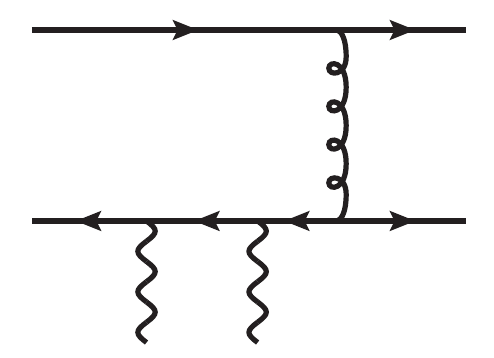} \\
		(B1) & (B2) & (B3) & (B4) & (B5) & (B6)
	\end{tabular}
	\caption{Hard scattering diagrams for the photon-proton scattering into a photon-pion pair. 
	The two incoming fermion lines on the left are from the diffracted nucleon, carrying momenta 
	$z_1 \hat{p}_1$ and $\bar{z}_1 \hat{p}_1 \equiv (1-z_1) \hat{p}_1$, respectively. 
	The two outgoing fermion lines on the right are to form the produced pion, carrying momenta 
	$z_2 \hat{q}_1$ and $\bar{z}_2 \hat{q}_1 \equiv (1-z_2) \hat{q}_1$, respectively. 
	The variables $z_1$ and $z_2$ are related to $x$ and $z$ by $z_1 = (x + \xi) / 2\xi$ and $z_2 = z$ (see the text).
	Another set of diagrams are also to be included by switching the two photon lines, giving 20 diagrams in total.}
	\label{fig:diagrams}
\end{figure}

The hard coefficients $\widetilde{C}_{\lambda\lambda'}$ ($C_{\lambda\lambda'}$) for the photon-proton scattering process are obtained from the diagrams in \fig{fig:diagrams} by amputating the parton lines associated with the diffracted proton and produced pion, and contracting them with 
$\gamma \cdot \hat{p}_1 / 2$ ($\gamma_5 \gamma \cdot \hat{p}_1 / 2$) and $\gamma_5 \gamma \cdot \hat{q}_1 / 2$, 
respectively, for the unpolarized (longitudinally polarized) GPD, 
where $\hat{p}_1 = (\Delta \cdot n) \bar{n}$ with $n=(0^+,1^-,{\bf 0}_T)$ and $\bar{n}=(1^+,0^-,{\bf 0}_T)$, 
and $\hat{q}_1 = (q_1 \cdot w)\bar{w}$ with $w$ and $\bar{w}$ like $n$ and $\bar{n}$, respectively, 
but with the corresponding $\hat{z}$-axis along $\vec{q}_{1}$. 
The helicity amplitudes are parametrized in terms of the center-of-mass energy squared $\hat{s} = s (1 + \xi) / (2\xi)$, 
the angles $(\theta, \phi)$ of the pion in the SDHEP frame, and the parton momentum fractions $x$ and $z$,
\begin{align} \label{eq:helicity-amplitudes}
	C_{\pm \pm}(x, z; \hat{s}, \theta, \phi) = \frac{\N}{\hat{s}} e^{\mp i \phi} C_+(x, z; \theta), 
	&\quad
	C_{\pm \mp}(x, z; \hat{s}, \theta, \phi) = \frac{\N}{\hat{s}} e^{\mp i \phi} C_-(x, z; \theta), 
	\nn\\
	\widetilde{C}_{\pm \pm}(x, z; \hat{s}, \theta, \phi) = \pm \frac{\N}{\hat{s}} e^{\mp i \phi} \widetilde{C}_{+}(x, z; \theta),
	&\quad
	\widetilde{C}_{\pm \mp}(x, z; \hat{s}, \theta, \phi) = \pm \frac{\N}{\hat{s}} e^{\mp i \phi} \widetilde{C}_{-}(x, z; \theta),
\end{align}
where 
$ \N = 2 i e^2 g^2 C_F / N_c $ is a normalization factor, 
and we have used parity symmetry to reduce the eight helicity amplitudes into four independent ones,
the two helicity-conserving ones, $C_+$ and $\wt{C}_+$, and two helicity-flipping ones, $C_-$ and $\wt{C}_-$.
To present these amplitudes with the charge-conjugation symmetry manifestly exhibited, we introduce the variables
$	z_1 = (x + \xi)/(2\xi) $
and
$z_2 = z$,
such that when we picture the parton pair state $A^* = [q\bar{q}']$ as a virtual meson with two valence partons of light-cone momentum fraction $z_1$ and $(1 - z_1)$, respectively.
Then the four independent helicity amplitudes are,
\begin{align}
&
2 \xi C_{+}(\theta; x, z) = 
	-(e_1-e_2)^2 
	\bb{
		\frac{1 - \cos \theta}{1 + \cos\theta} \cdot
			\P\frac{z_1 + z_2 - 2 z_1 z_2}{ 2 z_1 z_2 (1-z_1) (1-z_2) } 
	}
	+ (e_1^2 - e_2^2 ) 
	\bb{
		\frac{2}{1-\cos\theta} \cdot 
			\P\frac{z_1 - z_2}{z_1 z_2 (1 - z_1) (1 - z_2)}
	}
\nn\\
&\hspace{1.5em}
	+ e_1 e_2 \, \P 
	\bb{\frac{1 - \cos\theta }{z_1 z_2 (1-z_1) (1-z_2)} \cdot
		\frac{ \pp{ z_1 z_2 + (1 - z_1) (1 - z_2) } \pp{ z_1 (1 - z_1) + z_2(1 - z_2) }}{
			\pp{ 2 (1 - z_1)(1 - z_2) - (1 + \cos\theta) z_1 z_2 }
			\pp{ 2 z_1 z_2 - (1 + \cos\theta) (1 - z_1) (1 - z_2) } 
		}
	}
\nn\\
&\hspace{1.5em}
	+ i \pi \bigg\{
		(e_1-e_2)^2 \, \frac{2 \cos\theta}{\sin^2\theta}
			\pp{ \frac{ \delta(1 - z_1) }{z_2} + \frac{\delta(z_1)}{1 - z_2} }
		+ \frac{(e_1^2 - e_2^2)}{2} \cdot \frac{3 - \cos\theta}{1 - \cos\theta} 
			\pp{ \frac{\delta(1 - z_1)}{z_2} - \frac{\delta(z_1)}{1 - z_2} }
\nn\\
&\hspace{4em}
	- \frac{e_1 e_2 }{2} 
				\pp{ \frac{1 - \cos\theta}{1 + \cos\theta} - \frac{4}{1 - \cos\theta} }
   				\pp{ \frac{\delta(1 - z_1)}{z_2} + \frac{\delta(z_1)}{1 - z_2} }
			- \frac{e_1 e_2}{(1 + \cos\theta) \, z_2 \, (1 - z_2)} \times
\nn\\
&   
		\hspace{5.5em} \times 
		\bb{ 
			\pp{ \frac{z_1}{1 - z_2} + \frac{1 + \cos\theta}{2} \frac{1 - z_2}{z_1} }
				\delta\pp{ z_1 - \rho(z_2) }
			+ \pp{ \frac{1 + \cos\theta}{2} \frac{z_1}{1 - z_2} + \frac{1 - z_2}{z_1} }
				\delta\pp{ z_1 - \widetilde{\rho}(z_2) }
		}
	\bigg\} \,,
\label{eq:C+}
\\
&
2 \xi C_{-}(\theta; x, z) = 
	-(e_1 - e_2)^2
	\bb{ 
		\frac{1 - \cos\theta}{1 + \cos\theta} \cdot 
			\P\frac{z_1 z_2 + (1 - z_1)(1 - z_2)}{2 z_1 z_2 (1-z_1) (1-z_2)} 
	}
\label{eq:C-}
\\
&\hspace{1.5em}
	- i \pi \bigg\{
		\frac{(e_1 - e_2)^2}{1 + \cos\theta}  
			\pp{ \frac{\delta(1 - z_1)}{1 - z_2} + \frac{\delta(z_1)}{z_2} }
		+ \frac{e_1^2 - e_2^2}{2} 
			\pp{ \frac{\delta(z_1)}{z_2} - \frac{\delta(1 - z_1)}{1 - z_2} }
		+ \frac{2 e_1 e_2 }{1 + \cos\theta} 
			\pp{ \frac{\delta(1 - z_1)}{1 - z_2} + \frac{\delta(z_1)}{z_2} }
	\bigg\} \,,
\nn\\
&
2 \xi \widetilde{C}_{+}(\theta; x, z) =
	-(e_1 - e_2)^2 
		\bb{ 
			\frac{3 + \cos\theta}{ 2\,(1+\cos\theta) } 
				\cdot \P\frac{z_1 - z_2}{z_1 z_2 (1 - z_1) (1 - z_2)}
		}
\label{eq:Ct+}
\\
&	\hspace{1.5em}
	+ e_1 e_2 \, \P
	\bigg[ 
		\frac{( 3 + \cos\theta)}{z_1 z_2 (1-z_1) (1-z_2) } \cdot
			\frac{ (z_1 - z_2) (1 - z_1 - z_2)^2 }{ 
				\pp{ 2 (1 - z_1)(1 - z_2) - (1 + \cos\theta) z_1 z_2 } 
				\pp{ 2 z_1 z_2 - (1 + \cos\theta) (1 - z_1) (1 - z_2) }
			}
	\bigg]	
\nn\\
&\hspace{1.5em}
	+ i \pi \bigg\{
		\frac{2 (e_1 - e_2)^2}{\sin^2\theta} 
				\pp{ \frac{\delta(z_1)}{1 - z_2} - \frac{\delta (1 - z_1)}{z_2} }
		+ \frac{e_1^2 - e_2^2}{2} \, 
			\frac{1+ \cos\theta}{1- \cos\theta}
			\pp{ \frac{\delta(1 - z_1)}{z_2} + \frac{\delta(z_1)}{1 - z_2} }
		+ \frac{e_1 e_2}{2} \times
\nn\\
&\hspace{3.5em} \times
	\bb{ 
		\pp{ 
			\frac{8}{\sin^2\theta} - \frac{1 - \cos\theta}{1 + \cos\theta}
		}
		\pp{ 
			\frac{ \delta(z_1) }{1-z_2}  -  \frac{\delta (1 - z_1)}{z_2}
		} 
		- \frac{1- \cos\theta}{1 + \cos\theta} 
			\frac{z_1 - z_2}{z_2 \, (1 - z_2)}
			\bb{ 
				\delta \pp{ z_1 - \rho(z_2) } 
				+ \delta \pp{ z_1 - \widetilde{\rho}(z_2) }
			}
	}
	\bigg\} \,,
\nn\\
&
2 \xi \widetilde{C}_{-}(\theta; x, z) =
	-(e_1 - e_2)^2
		\bb{
			\frac{1 - \cos\theta }{1 + \cos\theta} \cdot
				\P \frac{(1 - z_1 - z_2)}{2 z_1 z_2 (1-z_1) (1-z_2)} 
		}
\label{eq:Ct-}
\\
&\hspace{1.5em}
	- i \pi \bigg\{
		\frac{(e_1 - e_2)^2}{1 + \cos\theta}  
			\pp{ \frac{\delta(z_1)}{z_2} - \frac{\delta(1 - z_1)}{1 - z_2} }
		+ \frac{e_1^2 - e_2^2}{2} 
			\pp{ \frac{\delta(z_1)}{z_2} + \frac{\delta(1 - z_1)}{1 - z_2} }
		+ \frac{2 \, e_1 e_2 }{1 + \cos\theta} 
			\pp{ \frac{\delta(z_1)}{z_2}  - \frac{\delta(1 - z_1)}{1 - z_2} }
	\bigg\} \,,
\nn
\end{align}
where $\P$ indicates that the hard coefficients should be understood in the sense of 
principle-value integration for $z_1$ (or $x$), when convoluted with the GPD and DA.
We have expressed the amplitudes in the general flavor case with the two parton lines 
carrying electric charge $e_1$ and $e_2$, with $e_u = 2/3$ and $e_d = -1/3$ for $u$ and $d$ quarks. 
For charged pion $\pi^{\pm}$ productions, we have $(e_1, e_2) = (e_u, e_d)$ or $(e_d, e_u)$, 
and all terms in Eqs.~\eqref{eq:C+}-\eqref{eq:Ct-} contribute.
For neutral pion production, however, we have $e_1 = e_2 = e_u$ or $e_d$, 
which cancels the terms proportional to $(e_1 - e_2)^2$ and $(e_1^2 - e_2^2)$.

The special gluon propagators in the type-A diagrams in \fig{fig:diagrams} introduce new poles of $z_1$ in addition to $0$ and $1$, 
\begin{align}
	\rho(z_2) = \frac{(1 + \cos\theta)(1 - z_2)}{1 + \cos\theta + (1 - \cos\theta) z_2}
		= \frac{1 - z_2}{1 + z_2 \tan^2(\theta/2)},
	\quad
	\widetilde{\rho}(z_2) = 1 - \rho(1 - z_2),
\end{align}
both of which lie between $0$ and $1$ for $z_2 \in [0, 1]$ and $\theta \in (0, \pi)$.
They translate to poles of $x$ at
\beq\label{eq:s7}
	x_p(\xi, z, \theta) = \xi \bb{ 2\rho(z) - 1 } 
		= \xi \cdot \bb{ \frac{\cos^2(\theta/2) (1-z) - z}{\cos^2(\theta/2) (1-z) + z} },
	\quad
	\wt{x}_p(\xi, z, \theta) 
		= \xi \bb{ 2\wt{\rho}(z) - 1 } = - x_p(\xi, 1-z, \theta),
\eeq
which lie between $-\xi$ and $\xi$.

The factorized helicity amplitudes are given by the convolution of the hard coefficients with the GPD $H$ or $\wt{H}$ and $\bar{D}(z)$, which  
can be simplified by using $\bar{D}(z) = \bar{D}(1-z)$.  
We have
\begin{align}
\M_+^{[\wt{H}]} 
	\equiv & \int_{-1}^1 dx \int_0^1 dz \, \wt{H}(x, \xi, t) \, \bar{D}(z) \, C_{+}(x, z; \theta)	\nn\\
	= & \pp{e_1-e_2}^2  \cdot \bar{D}_0 \cdot
			\bb{  
				\frac{1-\cos\theta }{2(1+\cos\theta) }\cdot \wt{H}^+_0(\xi, t)
				+ \frac{2 i \pi  \cos\theta}{\sin^2\theta} \cdot \wt{H}^+(\xi, \xi, t)
			}	\nn\\
	& + \pp{e_1^2-e_2^2} \cdot \bar{D}_0 \cdot 
			\bb{
				-\frac{2}{1-\cos\theta} \cdot \wt{H}^-_0(\xi, t)
				+ \frac{i \pi}{2} \cdot \frac{3-\cos\theta}{1-\cos\theta} \cdot \wt{H}^-(\xi, \xi, t)
			} 	\nn\\
	& + e_1 e_2 \cdot
			\cc{
				\int_0^1 \frac{\dd{z} \bar{D}(z)}{z(1-z)}
					\bb{ \frac{1}{2 z + (1 + \cos\theta) (1-z)} + \frac{2 z + (1 + \cos\theta) (1-z)}{2(1+\cos\theta)} } 
					\cdot \int_{-1}^{1} \dd{x} \frac{ \wt{H}^+(x, \xi, t) }{ x - x_p(\xi, z, \theta) + i \epsilon }
			\right.
			\nn\\
	&\hspace{4em}\left.
			+ \bar{D}_0 \cdot
			\bb{ \frac{1-\cos\theta}{2(1+\cos\theta)} \cdot \wt{H}^+_0(\xi, t)  
				- i\pi \pp{ \frac{1-\cos\theta}{2(1+\cos\theta)} - \frac{2}{1 - \cos\theta} } \cdot \wt{H}^+(\xi, \xi, t) 
			}
	}, 
	\label{eq:M+}\\
\M_-^{[\wt{H}]} 
	\equiv & \int_{-1}^1 dx \int_0^1 dz \, \wt{H}(x, \xi, t) \, \bar{D}(z) \, C_{-}(x, z; \theta)	\nn\\
	= & \pp{e_1-e_2}^2 \cdot \bar{D}_0 \cdot 
		\bb{
			\frac{1-\cos\theta}{2(1+\cos\theta)} \cdot \wt{H}^+_0(\xi, t)
			- \frac{i \pi}{1+\cos\theta} \cdot \wt{H}^+(\xi, \xi, t)
		}	\nn\\
	& + \pp{e_1^2-e_2^2} \cdot \frac{i \pi}{2} \cdot \bar{D}_0 \cdot \wt{H}^-(\xi, \xi, t)
		- e_1 e_2 \cdot \frac{2 i \pi }{1+\cos\theta} \cdot \bar{D}_0 \cdot \wt{H}^+(\xi, \xi, t) , 
	\label{eq:M-}\\
\Mt_+^{[H]} 
	\equiv & \int_{-1}^1 dx \int_0^1 dz \, H(x, \xi, t) \, \bar{D}(z) \, \wt{C}_{+}(x, z; \theta)	\nn\\
	= & \pp{e_1-e_2}^2  \cdot \bar{D}_0 \cdot 
		\bb{ 
			\frac{3 + \cos\theta}{ 2\,(1+\cos\theta) } \cdot H^+_0(\xi, t)
			- \frac{2 i \pi  }{\sin^2\theta} \cdot H^+(\xi, \xi, t)
		} 	\nn\\
	& + \pp{e_1^2-e_2^2} \cdot \frac{i \pi }{2}  
		\cdot  \frac{1+ \cos\theta}{1- \cos\theta} \cdot \bar{D}_0 \cdot H^-(\xi, \xi, t) 	\nn\\
	& + e_1 e_2 \cdot
		\cc{ 
			\int_0^1 \frac{\dd{z} \bar{D}(z)}{z(1-z)} 
			\bb{ \frac{1}{2 z + (1 + \cos\theta) (1-z)} - \frac{2 z + (1 + \cos\theta) (1-z)}{2(1+\cos\theta)} }
			\cdot \int_{-1}^{1} \dd{x} \frac{ H^+(x, \xi, t) }{x - x_p(\xi, z, \theta) + i \epsilon} 
			\right.
			\nn\\
	&\hspace{4em} \left.
			+ \bar{D}_0 \cdot
			\bb{
				\frac{3+\cos\theta}{2(1+\cos\theta)} \cdot H^+_0(\xi, t)
				- \frac{i \pi}{2} \pp{ \frac{8}{\sin^2\theta} - \frac{1-\cos\theta}{1+\cos\theta}}
					H^+(\xi, \xi, t)
			}
		}, 
	\label{eq:Mt+}\\
\Mt_-^{[H]} 
	\equiv & \int_{-1}^1 dx \int_0^1 dz \, H(x, \xi, t) \, \bar{D}(z) \, \wt{C}_{-}(x, z; \theta)	\nn\\
	= & -\pp{e_1-e_2}^2 \cdot \bar{D}_0 \cdot 
		\bb{
			\frac{1- \cos\theta }{2\, (1+ \cos\theta) } H^+_0(\xi, t) - \frac{i \pi }{1 + \cos\theta} H^+(\xi, \xi, t)
		}	\nn\\
	& - \pp{e_1^2-e_2^2} \cdot \frac{i \pi}{2} \cdot \bar{D}_0 \cdot H^-(\xi, \xi, t)
		+ e_1 e_2 \cdot \frac{2 i \pi }{1+\cos\theta} \cdot \bar{D}_0 \cdot H^+(\xi, \xi, t) ,
	\label{eq:Mt-}
\end{align}
where we have defined the charge-conjugation-even (C-even) and charge-conjugation-odd (C-odd) GPD combinations
\beq[eq:C-combination]
	H^{\pm}(x, \xi, t) \equiv H(x, \xi, t) \mp H(-x, \xi, t), \quad
	\wt{H}^{\pm}(x, \xi, t) \equiv \wt{H}(x, \xi, t) \pm \wt{H}(-x, \xi, t),
\eeq
and the ``zeroth moments'' of the DA and GPDs,
\beq
	\bar{D}_0 \equiv \int_0^1 \frac{dz \, \bar{D}(z)}{z}, \quad
	F_0(\xi, t) \equiv \P \int_{-1}^1 \frac{dx \, F(x, \xi, t)}{x - \xi},
\eeq
for $F$ being $H^{\pm}$ or $\wt{H}^{\pm}$. 
We note that charge conjugation symmetry on the hard coefficients are reflected as the symmetry under 
$(z_1, z_2) \leftrightarrow (1-z_1, 1-z_2)$ and $e_1 \leftrightarrow e_2$ in Eqs.~\eqref{eq:C+}-\eqref{eq:Ct-}. 
As a result, the $\pp{e_1-e_2}^2$ and $e_1 e_2$ terms are probing the C-even GPD components, whereas
the $\pp{e_1^2-e_2^2}$ terms the C-odd GPD components. 
Furthermore, in the $\pp{e_1-e_2}^2$ and $\pp{e_1^2-e_2^2}$ terms, the $\cos\theta$ dependence is decoupled from $z_1$ and $z_2$ dependence, which make the scattering amplitudes only sensitive to the moments and diagonal values of GPDs and vanish for shadow GPDs. In contrast, in the $e_1 e_2$ terms, the $\cos\theta$ dependence and the $z_1$ and $z_2$ dependence are entangled through $x_p(\xi,z,\theta)$ and $\widetilde{x}_p(\xi,z,\theta)$ in \eq{eq:s7}, which are capable of distinguishing shadow GPDs and provide 
enhanced sensitivity to the $x$ dependence of GPDs.

\subsection{B:\ Construction of shadow GPDs}
\label{app:shadow}

We define the (leading-order) shadow GPDs $S(x, \xi)$ by requiring
\beq[eq:sd gpd property]
	S(x, -\xi) = S(x, \xi),
	\quad
	S(\pm 1, \xi) = 0,
	\quad
	S(x, 0) = 0,
	\quad
	S(\pm \xi, \xi) = 0,
	\quad
	\int_{-1}^1 dx\, \frac{S(x, \xi)}{x - \xi} = 0,
\eeq
and the $(n+1)$-th moment of $S(x, \xi)$ to be an even polynomial of $\xi$ of at most $n$-th order,
\beq[eq:polynomiality]
	\int_{-1}^1 dx \, x^n \, S(x, \xi) = \sum_{i = 0, 2, \cdots}^n (2\xi)^i \, S_{n+1, i} \, ,
\eeq
which ensures the shadow GPDs to have the same polynomiality and time reversal properties as normal GPDs, while
they give null forward limits and vanishing contribution to the type of moment integral $\int dx\, {S}(x,\xi,t)/(x-\xi +i\varepsilon)$.
Note that we have dropped the $t$ dependence in $S$, which may be introduced
to relax the small $\xi$ suppression (due to $S(x, 0) = 0$) in \eq{eq:sd gpd property},
and the possible $\xi^{n+1}$ term in \eq{eq:polynomiality} which is associated with the $D$-term. 
We will construct a shadow $D$-term separately below.
Since it is either the C-even or C-odd GPD combination [\eq{eq:C-combination}] that enters the scattering amplitude,
we require $S(x, \xi)$ to be either odd or even in $x$, when it is to be added to $H$ or $\wt{H}$.
This has allowed us to leave out the condition $\int_{-1}^1 dx \, S(x, \xi) / (x + \xi) = 0$ in \eq{eq:sd gpd property}
from which it can be inferred.
Besides, we also require the first moment of the shadow GPD to vanish 
since that can be constrained by the electromagnetic form factor measurements,
i.e., 
\beq[eq:sgpd-1st-moment]
	\int_{-1}^1 dx \, S(x, \xi) = S_{1, 0} = 0.
\eeq

The conditions in \eq{eq:sd gpd property} lead to some general constraints on the shadow GPDs.
In low energy scattering such as at JLab Hall D, the accessible $\xi$ values are small, $\xi \ll 1$.
The zeros at $x = \pm \xi$ then severely constrain the shadow GPD values in the ERBL region, 
which can only grow up to a certain power of $\xi$.
In this case, the integrals in \eq{eq:sgpd-1st-moment} and the last equation in \eq{eq:sd gpd property} 
mainly receive contributions from the DGLAP region, which must be highly suppressed.
As a result, the shadow GPDs must have extra zeros in the DGLAP region, but not necessarily in the ERBL region.
To construct specific models for shadow GPDs, we choose the following ansatz,
\beq[eq:ansatz]
	S(x, \xi) = K_0 \, \xi^2 \, x^a \, (x^2 - \xi^2) \, (1 - x^2)^b \cdot Q_{2n}(x, c),
\eeq
where $a \geq 0$ and $b, n > 0$ are integers, and $Q_{2n}(x, c) = 1 + c \, x^2 + \cdots + q_{2n}(c) \, x^{2n}$ 
is an even $2n$-th order polynomial of $x$.
This parametrization automatically satisfies the first four conditions in \eq{eq:sd gpd property}.
Since it is only a fourth order polynomial of $\xi$, the polynomiality condition can be readily satisfied.
We have fixed the power of $(x^2 - \xi^2)$ to be unity; 
a higher power further suppresses the ERBL region and leads to an even smaller impact. 
For given $a$ and $b$, we choose $n$ to be the minimum integer such that \eq{eq:ansatz} satisfies all the conditions in 
Eqs.~\eqref{eq:sd gpd property}\eqref{eq:polynomiality} and \eqref{eq:sgpd-1st-moment}.
The single parameter $c$ is allowed to tune the shape of shadow GPD. 
For any given choice, we choose the normalization $K_0$ (independent of $\xi$) 
such that $\int_{-1}^1 dx \, S^2(x, \xi) = 2^2$ when $\xi = 0.1$.

We choose the GK model as the standard GPDs, $H_0(x, \xi, t)$ and $\wt{H}_0(x, \xi, t)$, and vary them by adding shadow GPDs
to the $u$ quark GPD.  
For the unpolarized GPD, we choose $a = 1$, $n = 3$, $b = 2$ or $6$, 
and $c = -11$ or $-17$ for two shadow GPDs, $S_1(x, \xi)$ and $S_2(x, \xi)$, to make up two new GPD models, $H_{1,2} = H_0 + S_{1,2}$.
Similarly, for the polarized GPD, we choose $a = 0$, $n = 3$, $b = 2$ or $6$, and $c = -24$ or $-40$ for 
two shadow GPDs, $\wt{S}_1(x, \xi)$ and $\wt{S}_2(x, \xi)$, to get two more GPD models, $\wt{H}_{1,2} = \wt{H}_0 + \wt{S}_{1,2}$.

For the unpolarized GPD, an additional term proportional to $\mathrm{mod}(n, 2) (2\xi)^{n+1}$ 
can exist on the right hand side of \eq{eq:polynomiality}, 
which comes from the $D$-term in the double distribution representation of GPD,
\beq
	H^q(x, \xi, t) = \int_{-1}^1 d\beta \int_{-1 + |\beta|}^{1 - |\beta|} d\alpha \, \delta(x - \beta - \xi \alpha) \, f^q(\beta, \alpha, t)
		+ \operatorname{sgn}(\xi) \, D^q(x / \xi, t) \, \theta\pp{ \xi^2 - x^2 } \,,
\eeq
where $D^q(x, t)$ is an odd function of $x$. 
Also, to retain the conditions in \eq{eq:sd gpd property}, we drop the $t$ dependence and choose the shadow $D$-term $D_s(x)$ 
such that
\beq[eq:ds property]
	D_s(-x) = - D_s(x),
	\quad
	D_s(1) = 0,
	\quad
	\int_{-1}^1 dx \, \frac{D_s(x)}{x - 1} = 0,
\eeq
where the subscript `s' is to remind that this $D$ term is to be part of the shadow GPD, 
but not to be the requirement for the $D$-terms in the normal GPDs.
Note that since the $D$-term automatically disappears in the forward limit, its magnitude does not necessarily suffer from 
the suppression when $\xi$ is small.
Because of the last condition in \eq{eq:ds property}, the shadow $D$-term cannot be 
probed by the dispersion relation in the DVCS data but it can modify the $D$-term in the gravitational form factor.
We choose the ansatz for the shadow $D$-term
\beq[eq:ds-ex]
	D_s(x) = J_0 \, x \, (1 - x^2) \cdot \pp{1 + c \, x^2 - \frac{7}{15}(3c + 5) x^4 } \, \theta(1 - x^2) \, ,
\eeq
with $c = 50$ and the normalization factor $J_0$ chosen to make $\int_{-1}^1 dx \, D_s^2(x) = 2^2$. 
Adding this to the $u$ quark GPD $H_0$ gives another GPD model, $H_3 = H_0 + D_s$.
